\documentclass[12pt]{iopart}

\def\mxth{\mathsurround=0pt }
\def\xversim#1#2{\lower2.pt\vbox{\baselineskip0pt \lineskip-.5pt
  \ialign{$\mxth#1\hfil##\hfil$\crcr#2\crcr\sim\crcr}}}             
\def\simgr{\mathrel{\mathpalette\xversim >}}                                    
\def\simle{\mathrel{\mathpalette\xversim <}}                                    
\usepackage{epsf,epsfig}

\newcommand{\be}{\begin{equation}}
\newcommand{\ee}{\end{equation}}
\newcommand{\bea}{\begin{eqnarray}}
\newcommand{\eea}{\end{eqnarray}}
\newcommand{\nn}{\nonumber}
\newcommand{\ba}{\begin{array}}
\newcommand{\ea}{\end{array}}
\renewcommand{\bi}{\begin{itemize}}
\newcommand{\ei}{\end{itemize}}

\newcommand{\N}{\chi^0}
\newcommand{\C}{\chi^\pm}

\newcommand{\sthto}{\ensuremath{SU(3) \x SU(2) \x U(1)}}                                                                      

\newcommand{\sto}{\ensuremath{SU(2) \x U(1)}}                                       
\newcommand{\x}{\ensuremath{\times}}                                                
\newcommand{\vect}[1]{\ensuremath{
\left( \begin{array}{c} #1    \end{array} \right) }}
\newcommand{\ra}{\ensuremath{\rightarrow}} 
\renewcommand{\slash}{\displaystyle{\not}}

\begin{document}
$\hspace{4.5in}\vcenter{\hbox{\bf MADPH-07-1479}
                \hbox{\bf hep-ph/0702001}
                \hbox{January 2007}}$

\title{TeV physics and the Planck scale}

\author{Vernon Barger$^1$, Paul Langacker$^2$, Gabe Shaughnessy$^1$}
\address{$^1$ Department of Physics, University of Wisconsin, Madison, WI 53706, USA }
\address{$^2$ School of Natural Sciences, Institute for Advanced Study, Einstein Drive, Princeton, NJ 08540, USA}

\begin{abstract}
Supersymmetry is one of the best motivated possibilities for new physics at the TeV scale.
However, both concrete string constructions and phenomenological considerations suggest the
possibility that the physics at the TeV scale could be more complicated than the
Minimal Supersymmetric Standard Model (MSSM), e.g., due to extended gauge symmetries,
new vector-like supermultiplets with non-standard \sto\ assignments,
and extended Higgs sectors. We briefly comment on some of these possibilities, and
discuss in more detail the class of extensions of the MSSM involving an additional
standard model singlet field. The latter provides a solution to the $\mu$ problem, and
allows significant modifications of the MSSM in the Higgs and neutralino sectors, with
important consequences for collider physics, cold dark matter, and electroweak baryogenesis.
\end{abstract}

\maketitle

\section{Beyond the standard paradigm}

There are many possibilities for the nature of new physics at the TeV scale,
but probably the most popular is the existence of TeV scale supersymmetry.
This is partly motivated by the elimination of quadratically divergent contributions
to the Higgs mass\footnote{Other possibilities include various forms of 
dynamical symmetry breaking~\cite{Hill:2002ap}, Little Higgs models~\cite{Perelstein:2005ka}, and large extra dimensions~\cite{Yao:2006px}.}, but other
rationales include a simple form of gauge coupling unification and the possible connection
to an underlying superstring theory and quantum gravity. Most work on supersymmetry or
discussions of the TeV scale from a superstring or grand unification point of view implicitly
assume most or all aspects a standard paradigm, which includes:

\begin{itemize}
\item The TeV scale is completely described by the 
Minimal Supersymmetric Standard Model (MSSM).
This leads to the possibility that the lightest supersymmetric particle (LSP), e.g., the lightest
neutralino, may be stable
and constitutes the cold dark matter. 
\item Supersymmetry is broken in a hidden sector which communicates only weakly
with our own sector (see, e.g., \cite{Chung:2003fi}, for possible mechanisms
for supersymmetry breaking and mediation).
Often some simple model for the soft supersymmetry breaking parameters is assumed,
such as minimal supergravity.
\item There may be an underlying  grand unified theory (GUT) \cite{Raby:2006sk}
broken at the unification scale $\sim 3\x 10^{16}$ GeV. This gives a simple explanation for
the observed supersymmetric gauge unification. 
\item There is a minimal seesaw model model for neutrino masses, involving very large Majorana masses
for the ``right-handed'' (\sto\ singlet) neutrinos. The out of equilibrium decays of the heavy neutrinos in the
early universe can lead (when combined with non-perturbative electroweak 
sphaleron effects) to the baryon asymmetry via the leptogenesis mechanism.
Specific seesaw models often assume GUT relations between Yukawa couplings, and often
invoke  large-dimensional Higgs
representations (e.g., the 120-plet of $SO(10)$).
\end{itemize}

However, there are possible phenomenological difficulties with this standard paradigm, and furthermore many of the ingredients may not easily emerge from an underlying string theory:

\begin{itemize}
\item Most of the problems of standard model (SM) remain, and new ones are introduced. In particular, the hierarchy of
the electroweak and Planck scales is stabilized but not explained, and there are possible
problems with flavor changing neutral
currents and electric dipole moments from loops involving supersymmetric partners.
 The MSSM also introduces the $\mu$ problem
problem (i.e., why the supersymmetric Higgs/Higgsino mass parameter $\mu$ in
the superpotential term   $W_\mu = \mu \hat H_u \cdot \hat H_d$ is comparable to
supersymmetry breaking) \cite{Kim:1983dt}. 
\item Most points in the landscape of string vacua with broken supersymmetry apparently lead to breaking
at a high scale, and in fact the possibility of high scale breaking has been considered
phenomenologically \cite{Arkani-Hamed:2004fb}.
\item It is likely that the soft supersymmetry breaking pattern is more complicated than
minimal supergravity \cite{Chung:2003fi}, leading to more challenging signals at the LHC 
\cite{Arkani-Hamed:2005px}.
\item It could well be that the new physics occurs primarily at the GUT or string scale, but
that remnants survive to the TeV scale. These would not necessarily solve any SM problems,
but be accidents of the underlying symmetry breaking and compactification. For example,
concrete superstring constructions (for recent constructions and reviews, see, e.g., 
\cite{stringconstruc}) often imply such new physics as
extended gauge groups, exotic fermions, and
extended Higgs/neutralino sectors\footnote{Other, more exotic,  possibilities include composite 
particles, charge $1/2$ particles (which may also be charged under a strongly coupled hidden sector group and confined),  other implications of a quasi-hidden sector, time-varying couplings, large extra dimensions,
and violation of the equivalence principle.}. These are often considered as defects of the
constructions, but perhaps they should be viewed as hints that the TeV scale physics is more complicated than
the MSSM.
\item It is difficult to generate viable 
four-dimensional GUT models from string compactifications 
\cite{stringconstruc}, so it is possible that any underlying
grand unification is actually broken in the higher-dimensional theory.
In that case, the canonical GUT Yukawa relations may not survive into the four-dimensional theory.
\item Even if one does have a GUT in the four-dimensional theory, the higher-dimensional
representations often invoked in model building (especially for neutrino masses) are unlikely or impossible
to emerge from string constructions.
Also, the enhanced constraints and symmetries from an underlying string theory may forbid couplings
needed for canonical bottom-up models (e.g., the right-handed neutrinos may be charged under
extended gauge symmetries)\footnote{There has been relatively little work on neutrino masses in
string constructions. Some of the existing classes of constructions do not have all of the couplings needed
for a canonical seesaw \cite{noss}, or lead to a seesaw that is not GUT-like \cite{nongut}.
Some constructions may generate neutrino masses via nonperturbative effects \cite{nonpert}, extended seesaws
\cite{ext}, Higgs triplets \cite{trip}, higher-dimensional operators \cite{hdo},  or small Dirac masses \cite{dir}.}.
\end{itemize}

For these reasons it is important to consider the possibility of  extensions of
(or alternatives to) the MSSM. We emphasize that such extensions do not necessarily solve problems of the
SM or lead to clean new signatures -- in many cases they make the physics at the LHC and other experiments 
or in cosmology more complicated and difficult. Nevertheless, they are a plausible and even
likely consequences of an underlying theory at the Planck or GUT scale.

\section{Extended gauge sectors}
Extensions of the standard model or of the MSSM frequently involve gauge groups
larger than \sthto. In particular, new $Z'$ gauge bosons associated with extra
$U(1)'$ factors are extremely common in string constructions~\cite{Cvetic:1995rj},
 dynamical symmetry breaking~\cite{Hill:2002ap}, Little Higgs models~\cite{Perelstein:2005ka},
and (as Kaluza-Klein excitations) in models with large extra dimensions~\cite{Yao:2006px}.
The generic reason is that if one starts with a higher gauge symmetry, it is typically
easier to break the non-abelian parts than  the $U(1)'$ factors.
In supersymmetric models both the electroweak and $U(1)'$-breaking scales are typically tied to the soft supersymmetry breaking scale~\cite{UMSSM}, so the $Z'$ mass is
usually expected to be a few TeV or less. The exception to this is when the $U(1)'$
breaking is associated with an $F$ and $D$-flat direction of the scalar potential\footnote{Grand
unified theories~\cite{Raby:2006sk}  with groups larger than $SU(5)$ also involve additional $U(1)'$s.
In this case, however, the $Z'$ mass is usually required to be at the GUT scale
so that additional particles which achieve their masses by $U(1)'$ breaking cannot mediate
rapid proton decay.}. Limits on the properties of a $Z'$ are model dependent, but for
electroweak-scale coupling to the quarks and leptons
one typically requires $M_{Z'} > 600-900$ GeV from Tevatron dilepton searches
as well as constraints from LEP 2 and weak neutral current experiments~\cite{Yao:2006px}. The $Z-Z'$
mixing angle must be
$ < { \rm few} \times 10^{-3}$  from $Z$-pole experiments. 
 If a heavy $Z'$ exists, it should be observable at the LHC and ILC for
 masses up to $\sim 5-8$ TeV through the resonant production of
 $e^+e^-,\ \mu^+ \mu^-$, or $q \bar q$, depending on the couplings and number of
 open channels into exotics and sparticles~\cite{Kang:2004bz}.
 For the lower end of this mass range (up to $1-2$ TeV) it should be possible to
 perform diagnostic probes of the $Z'$ couplings through 
 asymmetries, $y$ distributions, associated productions,  
 and  rare decays~\cite{diagnostics,Cvetic:1995zs,Leike:1998wr}.

The indirect implications of a heavy $Z'$ would be significant. For example,
the $U(1)'$ symmetry may forbid an elementary $\mu$ term, but allow a trilinear
superpotential coupling
\be \label{trilinear} W=\lambda \hat S \hat H_u \cdot \hat H_d,  \ee
where $S$ is a complex standard model singlet field charged under the $U(1)'$.
A non-zero expectation value $\langle S \rangle$ would not only give mass
to the $Z'$ but would also yield a dynamical effective $\mu$ parameter
\be
\label{mueff}
\mu_{\rm eff} = \lambda \langle S\rangle.
\ee
The $S$ scalar, the associated
neutralino, and the new $Z'$ gaugino have significant implications for collider
physics and cosmology, which are discussed for a wider class of models
in section \ref{xsinglet}. Other implications of a $U(1)'$ include
\begin{itemize}
\item Anomaly cancellation usually requires the introduction of new matter multiplets,
typically vector-like with respect to the SM group but charged under $U(1)'$~\cite{Erler:2000wu},
as discussed in the next section.
\item The $U(1)'$ may constrain the mechanisms for neutrino mass generation~\cite{Kang:2004ix}.
\item $Z'$ decays may be a significant source for the production of sparticles and exotics~\cite{Kang:2004bz,Baumgart:2006pa}.
\item String-motivated $Z'$ bosons often have family-nonuniversal charges,
leading to flavor changing neutral currents. For example, this could lead to
(small) tree-level effects in charmless $B$ decays that could compete with
enhanced loop effects in the large $\tan \beta$ limit of the 
MSSM~\cite{Langacker:2000ju,Leroux:2001fx,Barger:2003hg}.
\end{itemize}

\section{Extended matter sectors (quasi-chiral exotics)}
String constructions and other extensions often predict exotic new quarks
and leptons at the TeV scale. By exotic, we are referring to nonstandard
\sto\ representations, such as left-handed singlets or right-handed doublets\footnote{Much
more {\em exotic} exotics are also possible, such as color sextets,  or fractional or unusual
electric charges.}. We will assume for simplicity that such exotics
are vector-like with respect to \sthto\ because of strong constraints on new chiral particles
from precision electroweak (which are especially strong for an entire ordinary or mirror
family)~\cite{Yao:2006px}. However, they could be chiral with respect to an extended
gauge symmetry such as $U(1)'$, and in fact are usually needed for anomaly cancellation.
The effects of such exotics on gauge coupling unification are significant, unless
they are compensated, e.g.,  by embedding them in complete $SU(5)$ multiplets, allowing
high-scale supersymmetry breaking, or non-canonical $U(1)_Y$ normalization~\cite{Dienes:1996du,Barger:2007qb}. 

As an example, each family in the $E_6$ GUT model~\cite{Barger:1985nq,Hewett:1988xc,King:2005jy} is contained in a 27-plet,
which contains, in addition, two standard model singlets (candidates for the $S$ fields
described in the previous section and right handed neutrinos);
a vector-like pair of heavy charge $-1/3$ $SU(2)$-singlet quarks  $D_L + D_R$;
and a vector-like $SU(2)$-doublet pair  $\vect{E^0 \\ E^-}_L$+ $\vect{E^0 \\ E^-}_R$ 
of exotic leptons or Higgs fields. These exotic states are chiral with respect to the
additional $U(1)'$ symmetries in $E_6$\footnote{The $E_6$ model is a convenient
example of anomaly-free quantum numbers and $U(1)'$ charges. However, 
for TeV-scale $U(1)'$ breaking the $E_6$ Yukawa relations would need to be 
broken to forbid rapid proton decay.}.

Let us focus on the example of the exotic $D$ quarks, which can be pair-produced
by QCD processes at a hadron collider, and their scalar supersymmetric partners
$\tilde D$, produced with an order of magnitude smaller cross section.
(The rates are of course smaller for exotic leptons). Once produced, there
are three major decay possibilities for  $D$ or $\tilde D$~\cite{Barger:1985nq}-\cite{kln}:
\begin{itemize}
\item The decay may be $D\ra u_iW^-, \ D \ra d_i Z$, or $ D \ra d_i H^0$, if driven by  mixing
with a light charge $-1/3$ quark. The current limit is  $m_D \simgr$ 200 GeV~\cite{Andre:2003wc},
which should be improved to $\sim$ 1 TeV at the LHC. Note, however, that 
such mixing is forbidden in the supersymmetric $E_6$ model if $R$-parity is 
conserved~\cite{kln}.
\item One may have $\tilde D$ \ra \ quark jets if there is a small diquark operator $ \bar u \bar u \bar D$, 
or $\tilde D$ \ra \ quark jet $+$ lepton for a leptoquark operator $ l q  \bar D$.
Such operators do not by themselves violate $R$ parity, and therefore allow a stable lightest
supersymmetric particle. They are strongly constrained by the $K_L-K_S$ mass difference
and by $\mu-e$ conversion, but may still be significant~\cite{kln}.
\item They may be stable at the renormalizable level due to an accidental symmetry (e.g., from an
extended gauge group), so that they hadronize and escape from or stop in the detector~\cite{kln},
with signatures~\cite{Kraan:2005ji} somewhat similar to the quasi-stable gluino
expected in split supersymmetry~\cite{Arkani-Hamed:2004fb}. They could then decay
by higher-dimensional operators on a time scale of $\simle 100$ s, short enough to
avoid cosmological problems~\cite{Kawasaki:2004qu}.
\end{itemize}

\section{Extended Higgs sectors}
String constructions frequently lead to extended Higgs sectors. In particular, models predicting
additional pairs of light Higgs doublets $H_{u,d}$ are very common.
Supersymmetric models with more than a single pair of Higgs doublets have not been much studied,
in part because they make the physics more complicated without necessarily leading to
distinctive signatures. Clearly, they would lead to a richer Higgs/Higgsino spectrum and
decay possibilities, and would expand the possibilities for models of fermion masses and mixings
(e.g., if additional symmetries or constraints from the underlying theory restrict the allowed
couplings of each of the Higgs doublets). The extra neutral Higgs fields could lead
to flavor changing neutral currents and CP violating effects (though suppressed by Yukawa
couplings). They would also significantly modify gauge coupling unification unless accompanied 
by other exotic supermultiplets (e.g., to form complete $SU(5)$-like multiplets).

Another possibility involves extensions of the MSSM with new standard model singlet fields $S_i$,
which may, however, be charged under additional $U(1)'$ or other symmetries. 
The existence of such singlets leads to relaxed upper {\em and} lower limits on the
lightest Higgs compared to the MSSM, modifies the spectrum and production and
decay prospects for the Higgs and neutralino particles, 
expands the MSSM allowed parameter range (e.g., 
$\tan \beta=v_u/v_d$ can be close to 1), and modifies the possibilities for cold dark matter
and electroweak baryogenesis.
In the next section, we
discuss some of the implications of these singlet-extended models in more detail.

\section{Singlet extensions of the MSSM}
\label{xsinglet}

Supersymmetric models with an additional singlet Higgs field address the $\mu$ problem by promoting the $\mu$ parameter to a dynamical field whose vacuum expectation value $\langle S\rangle$ and coupling $\lambda$ determine the effective $\mu$-parameter $\mu_{\rm eff}$,
as in the $U(1)'$ example in (\ref{mueff}).
Depending on the symmetry imposed on the theory to forbid an elementary $\mu$, a variety of singlet extended models (xMSSM) may be realized: see Table \ref{tbl:models}.  The models we focus on include the Next-to-Minimal Supersymmetric SM (NMSSM) \cite{NMSSM}, the Nearly-Minimal Supersymmetric SM (nMSSM) \cite{nMSSM,Dedes:2000jp,Menon:2004wv}, and the $U(1)'$-extended MSSM (UMSSM) \cite{UMSSM}, as detailed in Table \ref{tbl:models} with the respective symmetries.  A Secluded $U(1)'$-extended MSSM (sMSSM) \cite{sMSSM,Han:2004yd}, motivated by some string
constructions, contains three singlets in addition to the standard UMSSM Higgs singlet.
The additional singlets allow a large (TeV scale) $Z'$ mass and a smaller $\mu_{\rm eff}$. This model is equivalent to the nMSSM in the limit that the additional singlet vevs are large, and the trilinear singlet coupling, $\lambda_s$, is small \cite{Barger:2006dh}.   The nMSSM and sMSSM will therefore be referred to together as the n/sMSSM.  The additional singlet state of the extended models gives additional Higgs bosons and neutralino states.  The number of Higgs and neutralino states in the various models are summarized in Table \ref{tbl:models}.

\begin{table}[ht]
\begin{tabular}{|c|ccccc|}
\hline
Model:& MSSM &NMSSM &nMSSM& UMSSM & sMSSM\\
\hline
Symmetry:  &--  &~~ $Z_3$    & $Z^R_5, Z^R_7$      & ~~$U(1)'$&$U(1)'$ \\\hline
Extra   &--         &       ~~${\kappa\over3} \hat S^3$    & $t_F  \hat S$& ~~-- &$\lambda_S  \hat S_1  \hat S_2  \hat S_3$ \\
superpotential term&   --      &      (cubic)    & (tadpole) & ~~-- & (trilinear secluded)\\\hline
 $\N_i$ &        	4	  &	~~~5	&  	5	 & ~~6 & 9\\
 $H^0_i$ &       	2	  &	~~~3	&  	3	 & ~~3 & 6\\
 $A^0_i$ &        	1	  &	~~~2	&  	2	 & ~~1 & 4\\
\hline
\end{tabular}
\caption{Symmetries associated with each model and their respective terms in the superpotential; the number of states in the neutralino ($\chi^0_i$)  and neutral Higgs sectors (CP-even: $H_i^0$; CP-odd: $A_i^0$) are also given.  All models have two charginos, $\C_i$, and one charged Higgs boson, $H^\pm$.  We ignore possible CP violation in the Higgs sector.}
\label{tbl:models}
\end{table}

The additional CP-even and CP-odd Higgs boson, associated with the inclusion of the singlet field, yield interesting experimental consequences at colliders.  For recent reviews of these models including their typical Higgs mass spectra and dominant decay modes, see Ref. \cite{Barger:2006dh,Kraml:2006ga}.

The tree-level Higgs mass-squared matrices are found from the potential, $V$, which is a sum of the $F$-term, $D$-term and soft-terms in the lagrangian, as follows.
\bea
V_F &=& |\lambda H_u\cdot H_d+t_F+ \kappa S^2|^2 + |\lambda S|^2 \left(|H_d|^2+|H_u|^2 \right), \\
V_D &=& \frac{G^2}{8}\left( |H_d|^2-|H_u|^2 \right)^2+ \frac{g_{2}^2}{2} \left( |H_d|^2|H_u|^2-|H_u \cdot H_d|^2 \right),\\
 &+& {{g_{1'}}^2\over2}\left(Q_{H_d} |H_d|^2+Q_{H_u} |H_u|^2+Q_{S} |S|^2\right)^2\\ \nn
V_{\rm soft}&=&m_{d}^{2}|H_d|^2 + m_{u}^{2}|H_u|^2+ m_s^{2}|S|^2 + \left( A_s \lambda S H_u\cdot H_d + {\kappa \over 3} A_{\kappa} S^3+t_S  S + h.c. \right).
\label{eq:potential}
\eea
Here, the two Higgs doublets with hypercharge $Y=-1/2$ and $Y=+1/2$, respectively, are
\be
H_d = \left( \begin{array}{c} H_d^0 \\ H^- \end{array} \right), \qquad
H_u = \left( \begin{array}{c} H^+ \\ H_u^0 \end{array} \right).
\ee
and $H_u \cdot H_d = \epsilon_{ij} H_u^i H_d^j$.  For a particular model, the parameters in $V$ are understood to be turned-off appropriately
\bea
{\rm NMSSM}&:& g_{1'}=0,\  t_F = 0,\  t_S = 0,\nn\\
{\rm nMSSM}&:& g_{1'}=0,\ \kappa=0,\ A_\kappa = 0, \\
{\rm UMSSM}&:& t_F = 0,\ t_S = 0,\ \kappa = 0,\ A_\kappa = 0.\nn
\eea
The couplings $g_1,g_2$, and ${g_{1'}}$ are for the $U(1)_Y$, $SU(2)_L$, and $U(1)'$ gauge symmetries, respectively, and the parameter $G$ is defined as $G^2=g_1^2+g_2^2$.  
The NMSSM model-dependent parameters are $\kappa$ and $A_\kappa$ while the nMSSM parameters are $t_F$ and $t_S$.  The model dependence of the UMSSM is expressed by the $D$-term that has the $U(1)'$ charges of the Higgs fields, $Q_{H_d}, Q_{H_u}$ and $Q_S$ and can be expressed in terms of the $E_6$ breaking angle:

\bea
Q_{H_d}&=& -{1\over \sqrt 10} \cos \theta_{E_6} - {1\over \sqrt 6} \sin \theta_{E_6}\nn\\
Q_{H_u}&=& {1\over \sqrt 10} \cos \theta_{E_6} - {1\over \sqrt 6} \sin \theta_{E_6} \\
Q_S &=& - Q_{H_d}-Q_{H_u}\nn
\eea

One loop radiative corrections to the Higgs mass can be large due to the large top quark Yukawa coupling.  At the one-loop level, the top and stop loops are the dominant contributions.  Gauge couplings in the UMSSM are small compared to the top quark Yukawa coupling so the one-loop gauge contributions can be dropped.  The model-dependent contributions do not affect the Higgs mass significantly at one-loop order.  Thus, the usual one-loop SUSY top and stop loops are universal in these models.  The one-loop corrections to the potential are derived from the Coleman-Weinberg potential.

\subsection{Particle spectra}

\begin{figure}[t]
\hspace{.5in}
\includegraphics[angle=-90,width=0.49\textwidth]{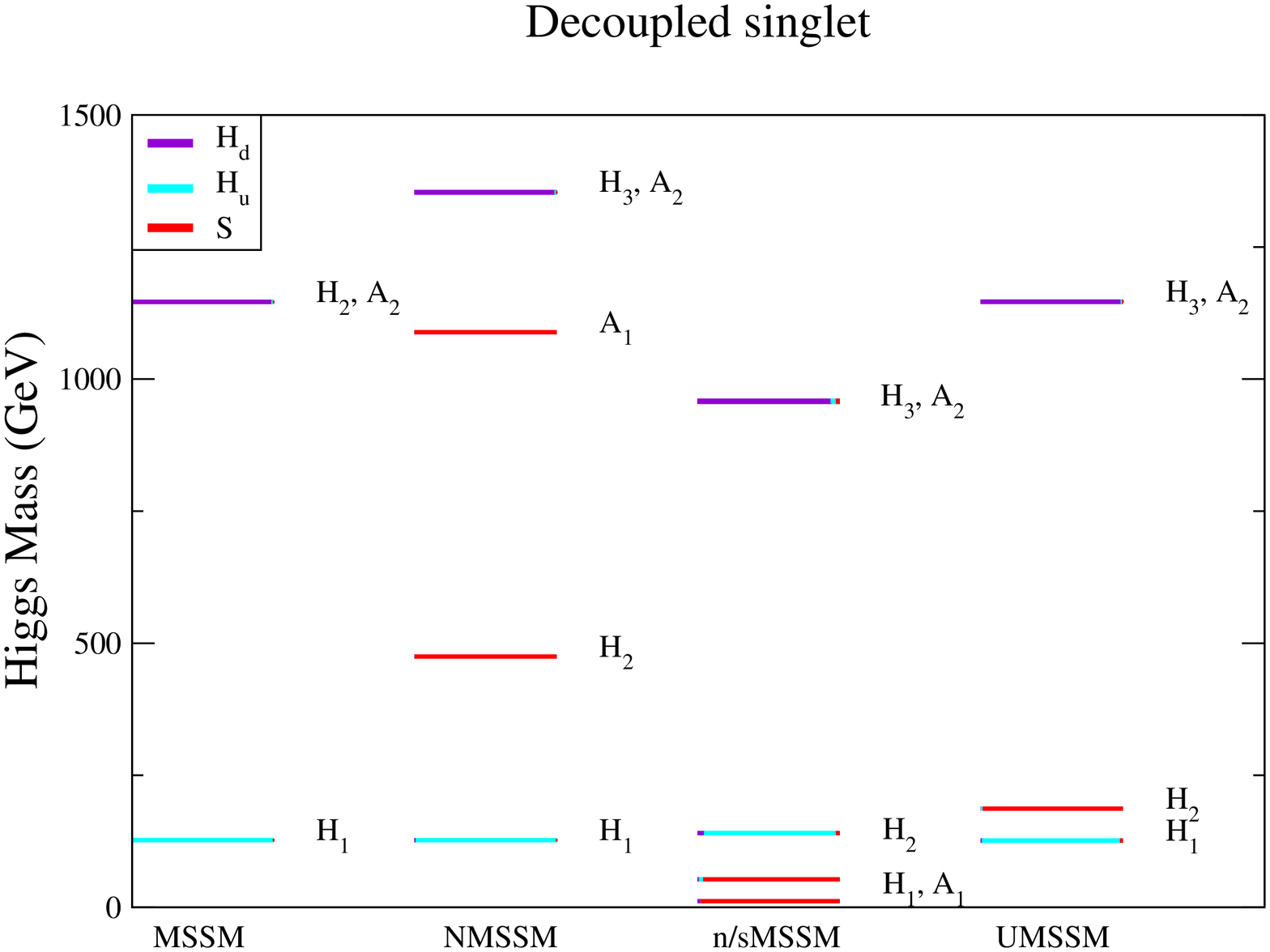}
\includegraphics[angle=-90,width=0.49\textwidth]{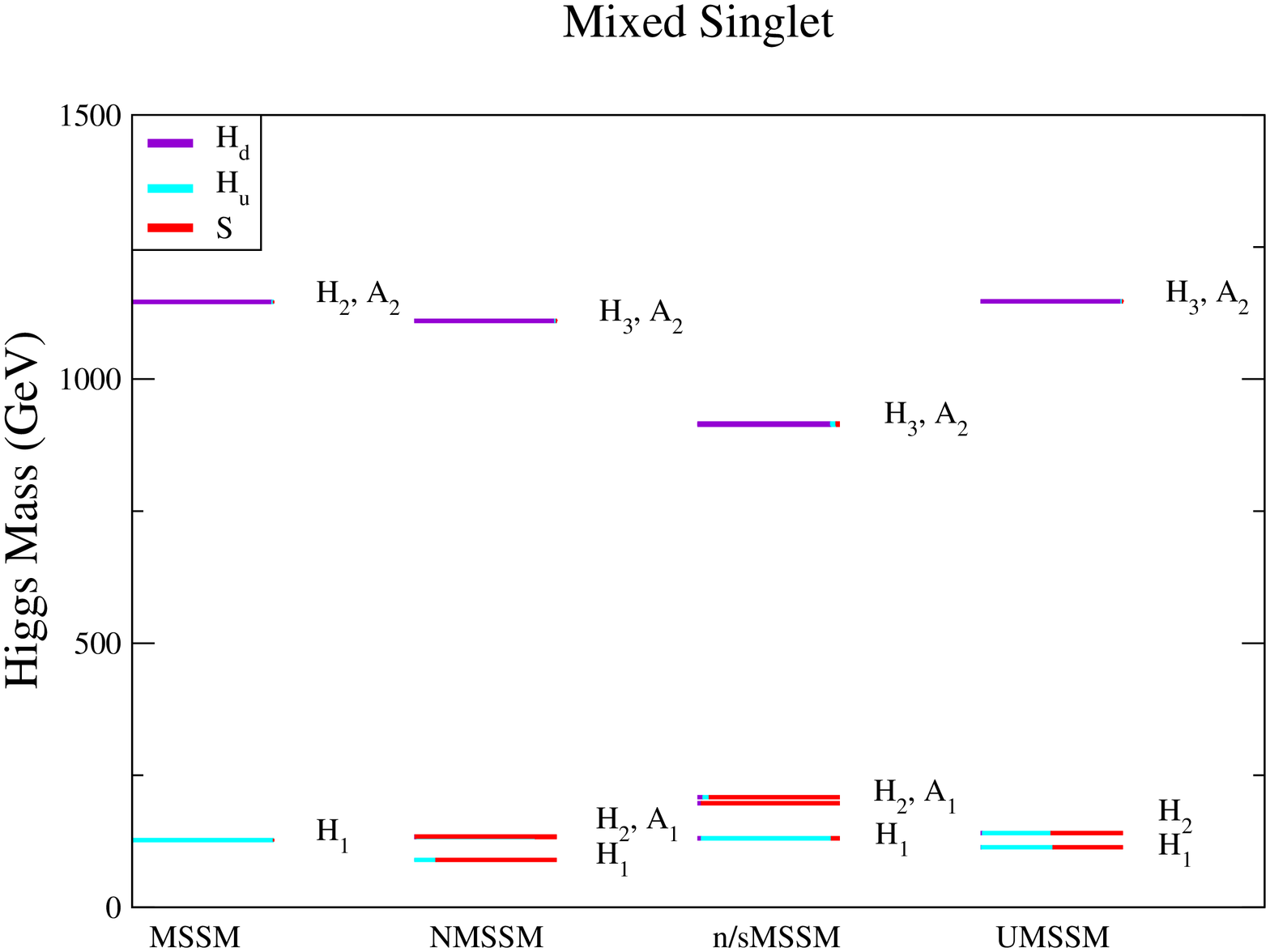}
\caption{Illustrative Higgs composition $(H_d, H_u, S)$ for the models in (a) a decoupled singlet scenario and (b) a strongly mixed singlet scenario.  In the decoupled scenario, the extended model has a spectrum similar to that of the MSSM, but contains an additional singlet Higgs that is heavy in the NMSSM and UMSSM and light in the n/sMSSM.  Figures from Ref. \cite{Barger:2006new}. }
\label{fig:illust}
\end{figure}

To illustrate the Higgs sector of the extended models in the cases in which the additional Higgs is either decoupled or strongly mixed with the MSSM Higgs boson, we present in Figure \ref{fig:illust} the neutral Higgs mass spectra for particular points in parameter space.

\begin{figure}[htbp]
\begin{center}
\includegraphics[angle=-90,width=0.49\textwidth]{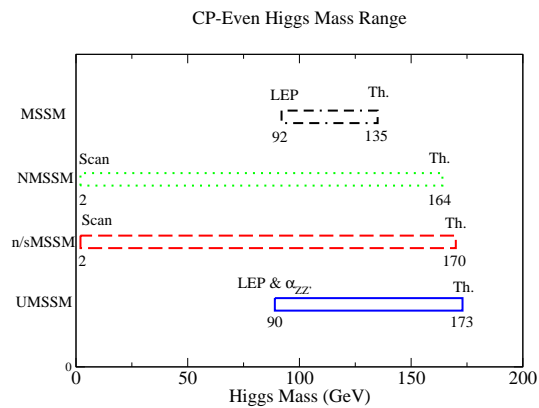}
\caption{Lightest CP-even Higgs mass range.  Figure from Ref. \cite{Barger:2006dh}.}
\label{fig:massrange}
\end{center}
\end{figure}

With sufficient mixing, the lightest Higgs boson can evade the current LEP bound~\cite{Sopczak:2006vn} on the SM Higgs mass in these models \cite{Barger:2006dh}.  Alternatively, singlet interactions increase the lightest Higgs mass-squared by ${\cal O}(\frac{1}{2} \lambda^2 v^2\sin^2 2 \beta)$, allowing it to be in the theoretically excluded region in the MSSM for low $\tan \beta$ \footnote{Additional gauge interactions contribute to this increase with size ${\cal O}(g_{1'}^2 v^2(Q_{H_u}^2 \cos^2\beta+Q_{H_d}^2 \sin^2 \beta))$ in the UMSSM.}.  The lightest Higgs mass ranges for each model are shown in Figure \ref{fig:massrange}.

A light decoupled $H_1$ is often accompanied by a MSSM-like $H_2$ with a mass in the 115-135 GeV range and MSSM strength couplings to SM fields.  

\begin{figure}[t]
\hspace{.5in}
\includegraphics[angle=-90,width=0.49\textwidth]{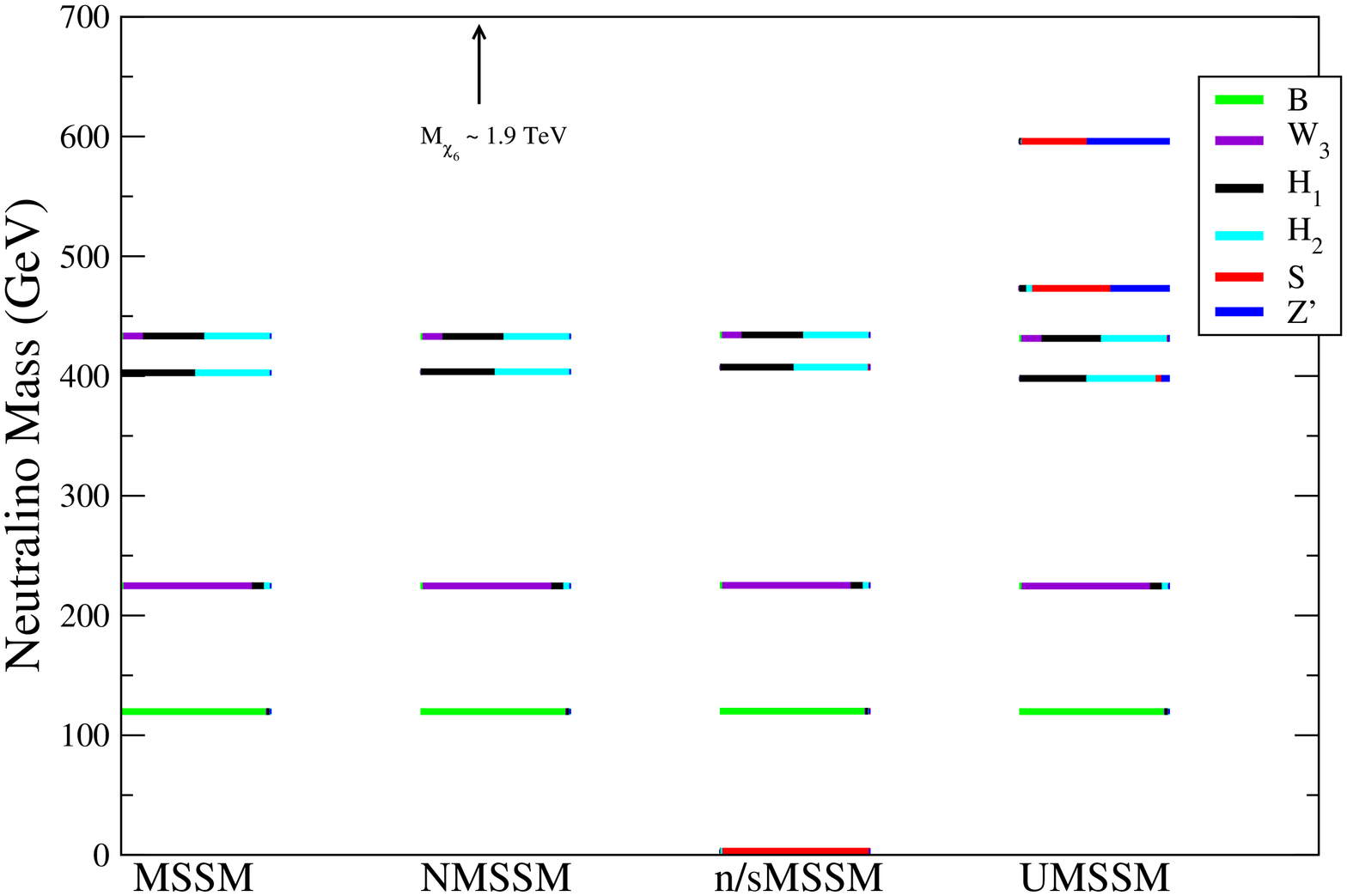}
\includegraphics[angle=-90,width=0.49\textwidth]{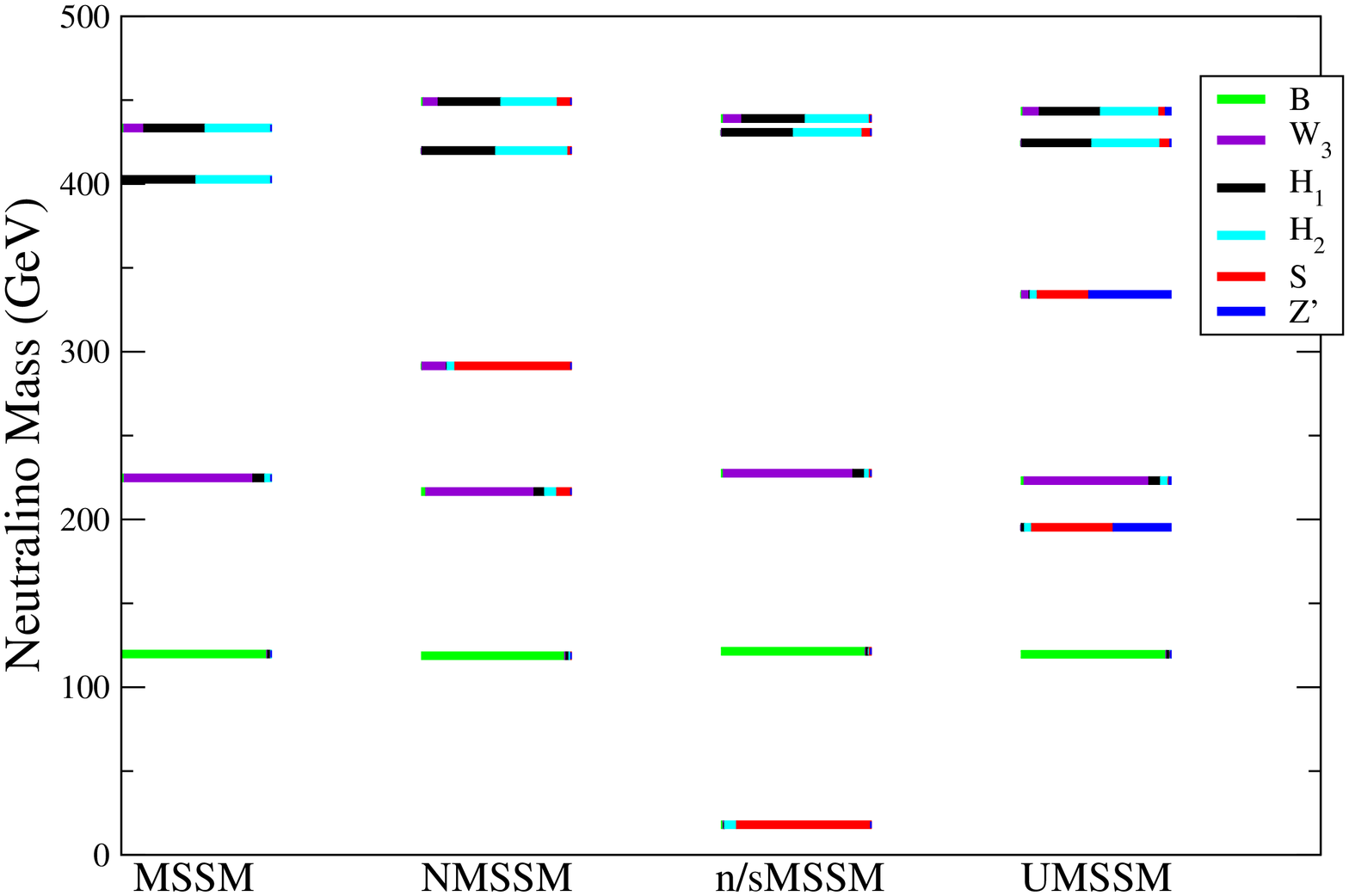}
\caption{Illustrative neutralino composition for the models in (a) a decoupled singlino scenario and (b) a strongly mixed singlino scenario.  Here, the MSSM contains a light Bino and Wino and heavy Higgsinos.  The NMSSM has a similar spectrum, but contains an additional heavy neutralino, while the n/sMSSM has a very light extra neutralino.  The UMSSM has two additional neutralinos that can intermix; their masses are strongly dependent on the singlet Higgs charge under the $U(1)'$ symmetry and the corresponding gaugino mass value.  Figures from Ref. \cite{Barger:2006kt}.}
\label{fig:level-light}
\end{figure}

At least one new neutralino state beyond the MSSM exists, the superpartner of the singlet, the singlino.  If a $U(1)'$ gauge symmetry exists, the $Z'$ino increases the neutralino states to six.  The neutralino states can include four MSSM-like states and one nearly decoupled singlino state, or the singlino can significantly mix with the other states, as determined from the neutralino mass matrix
\be
{\cal M}_{\N} = \left( \begin{array} {c c c c |c| c}
	M_1 	&	0&	{-g_1 v_d/ 2}&	{g_1 v_u / 2}&	0&  0\\
	0 	&M_2&	{g_2 v_d / 2}&	{-g_2 v_u / 2}&	0&  0\\
	{-g_1 v_d / 2} 	&	{g_2 v_d / 2}&	0&	-\mu_{\rm eff}&	-\mu_{\rm eff}v_u/s&  {g_{1'}} Q_{H_d} v_d\\
	{g_1 v_u / 2} 	&	{-g_2 v_u / 2}&	-\mu_{\rm eff}& 0&	-\mu_{\rm eff}v_d/s&  {g_{1'}} Q_{H_u} v_u\\
	\hline
	0&0&-\mu_{\rm eff} v_u/s&-\mu_{\rm eff} v_d/s&\sqrt 2 \kappa s&{g_{1'}} Q_{S} s\\
	\hline
	0&0&{g_{1'}} Q_{H_d} v_d&{g_{1'}} Q_{H_u} v_u&{g_{1'}} Q_{S} s& M_{1'}\\
	\end{array} \right),
	\label{eq:neutmass}
\ee
where $M_1$, $M_2$ and $M_{1'}$ are the gaugino masses of the $U(1)$, $SU(2)$ and $U(1)'$ gauge symmetries. The fifth (sixth) rows and columns refer to the singlino ($Z'$ino).  Gaugino mass unification is assumed, constraining $M_{1'}=M_1={5 g_1^2\over3 g_2^2}M_2$ at low scales.  The resulting neutralino spectrum can be substantially altered with respect to the MSSM.  Figure \ref{fig:level-light} illustrates the neutralino spectrum and composition for a decoupled and mixed scenario of singlino (and $Z'$ino for the UMSSM) mixing.  

The neutralino in the n/sMSSM is very light, often below 50 GeV.  A very light neutralino in the n/sMSSM  has important implications for cosmology and dark matter direct detection, see Section \ref{sect:dm}.  A very light singlino is less natural but can also be achieved in the NMSSM with a very small value of $\kappa$, as the $\kappa \to 0$ limit corresponds to the n/sMSSM.  The lightest neutralino in the UMSSM is typically MSSM-like, but can be dominantly singlino and $Z'$ino.

\subsection{Collider signatures}

Singlet mixing can strongly affect the observation of the Higgs boson at the LHC.  The branching fractions for discovery channels of the Higgs boson in the SM can be suppressed significantly.  Since the couplings to gauge bosons are at most SM strength, production rates are usually smaller than in the SM.

The most promising discovery channel over most of the Higgs mass range is the golden channel $H_i \to ZZ^* \to 4l$, since it has very low backgrounds.  This channel is expected to permit SM Higgs discovery for Higgs masses $120- 600$ GeV.  In extended models the signal is reduced by a factor of $\xi^2_{VVH_i}\times {Bf(H\to ZZ)\over Bf(h_{SM}\to ZZ)}$ compared to the SM, where $\xi_{VVH_i}$ is the $VVH_i$ coupling relative to the SM.  Therefore, it is possible that the Higgs in the extended models is missed via direct searches.

For light Higgs bosons ($m_H < 120$ GeV) the decay $H\to \gamma \gamma$ has the best significance.  Combining this mode with $H\to ZZ\to 4l$ yields a total significance above $5\sigma$ required for discovery for the lightest Higgs boson in the SM.  For some parameter points, the decay $H\to \gamma \gamma$ is enhanced in the extended models due to a larger Yukawa coupling or interference effects \cite{Barger:2006dh}. The Higgs production and decay, and detection possibilities at the LHC, are
further discussed in~\cite{Barger:2006dh,Barger:2006new}.

Due to the shifts in the neutralino spectrum compared to the MSSM, the cascade decay chains may be substantially modified \cite{Barger:2006kt,Ellwanger:1997jj}.  In particular, multilepton events such as a 5 lepton or 7 lepton signature are possible.  Chargino decays are indirectly affected via their decays to a lighter neutralino state.  The number of neutralino states lighter than the chargino and their modified compositions alter the chargino branching fractions.  This is typically found in the n/sMSSM, where the chargino can decay to an MSSM like $\N_2$ and a singlino $\N_1$, yielding a 5 lepton signal.  Additionally, the extra step in a neutralino decay can allow a 7 lepton final state.  Other models can also exhibit this behavior, but less naturally.

In some cases the neutralino can be light enough to spoil the chances for direct Higgs discovery.  The Higgs boson may have a dominant invisible decay to stable neutralinos that are undetected except as missing transverse energy, $\slash{E}_T$.  When the $H\to \N_1 \N_1$ decay channel  is open, the Higgs is generally invisible \footnote{There are some corners of parameter space which allow $H_1\to A_1 A_1$ with the $A_1$ mass below the threshold for decays to bottom pairs \cite{Dermisek:2005ar}.}.  This invisible decay is usually kinematically inaccessible for the MSSM, NMSSM, and UMSSM due to the lower limit on $m_{\N_1}$ of 53 GeV which is correlated with the lower bound of the chargino mass, $M_{\chi_1^\pm} > 104$ GeV~\cite{Barger:2006dh}.  

Invisible decays are often dominant in the n/sMSSM where the lightest neutralino mass is typically lighter than 50 GeV \cite{Menon:2004wv,Barger:2006dh,Barger:2006kt,Barger:2005hb}.  Therefore, traditional searches for the discovery of $H_1$ are unlikely for some parameter regions of the n/sMSSM.  However, indirect discovery of an invisibly decaying Higgs is possible in weak boson fusion and in $Z$-Higgstrahlung at the LHC \cite{Eboli:2000ze,Davoudiasl:2004aj} with jet azimuthal correlations and $p_T$ distributions or via the $Z$ recoil spectrum at the ILC.

\subsection{Cosmological dark matter and recoil detection}
\label{sect:dm}

\begin{figure}[tb]
\hspace{.7in}
\includegraphics[width=0.35\textwidth,angle=-90]{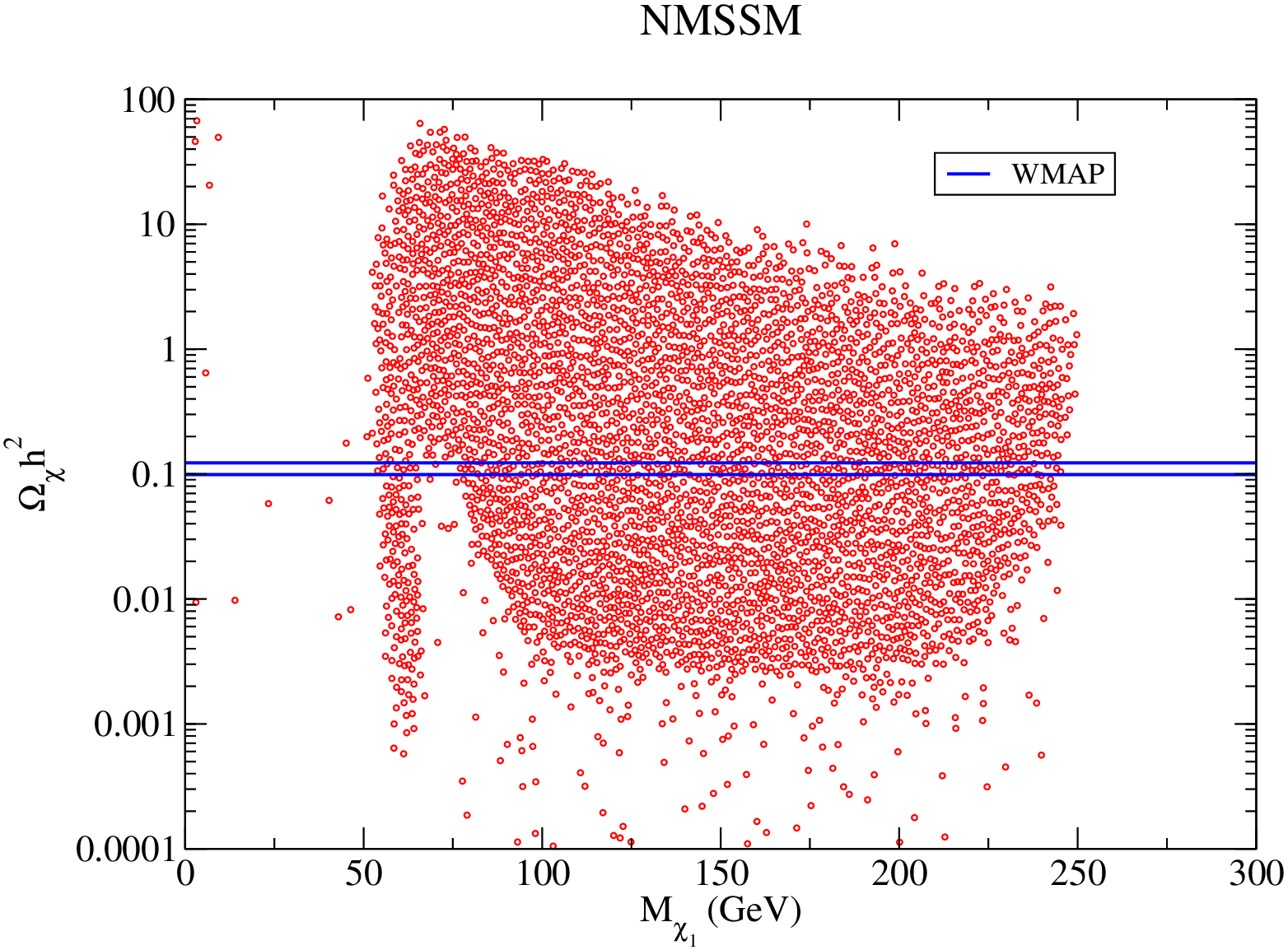}
\includegraphics[width=0.35\textwidth,angle=-90]{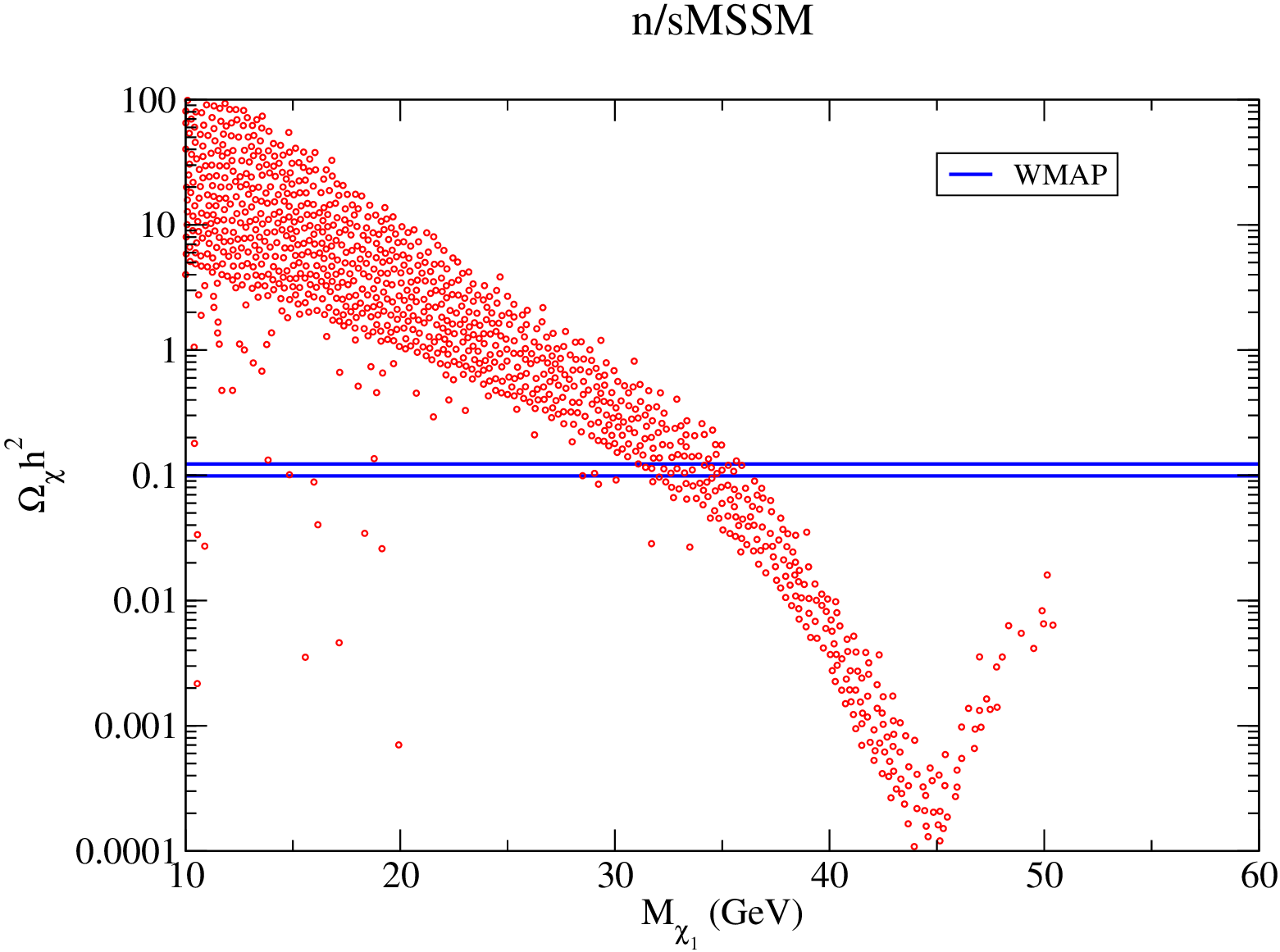}
\caption{Neutralino relic density versus the lightest neutralino mass.  The NMSSM distribution of points is very similar to the MSSM and UMSSM.  The relic density is constrained to be in the region $0.123>\Omega h^2>0.099$ provided that the model (with thermal production) is solely responsible for the observed dark matter.  The efficient annihilations through the Higgs boson pole in the MSSM, NMSSM and UMSSM are evident at $m_{\N_1} \sim M_{H_1}/2 \sim 60$ GeV and through the $Z$ boson pole at $m_{\N_1}\sim M_Z / 2$ in the n/sMSSM. Figures from Ref. \cite{ds-xmssm}.}
\label{fig:rd}
\end{figure}

The lightest neutralino in these models is expected to be the source of the relic abundance of dark matter in the universe.  The relic abundance observed by WMAP and other experiments  places constraints on these models, specifically the n/sMSSM \cite{Spergel:2006hy}.  In Figure \ref{fig:rd}, we show the thermal relic density in the NMSSM and n/sMSSM along with the $2\sigma$ uncertainty band of the observed relic density.  $\Omega_\chi$ is the dark matter density in units of the critical density for a closed universe and $h$ is the Hubble constant ($h=0.72\pm 0.08$ from the HST measurement \cite{Freedman:2000cf}).

The distribution of points in the MSSM, NMSSM and UMSSM are similar.  The mass of the lightest neutralino for allowed points in the n/sMSSM is typically $> 30$ GeV which is dominated by the $s$-channel $Z$ boson exchange.  However, lighter masses may fulfill the constraints by annihilation through very light CP-odd Higgs bosons or through further decays into the additional neutralinos in the sMSSM \cite{ds-xmssm}.  The dark matter constraints are less restrictive for the other models, with most of the neutralino mass range satisfying the WMAP observation.  The points within the observed $\Omega_\chi h^2$ range for masses above $m_{\N_1}\sim 80$ GeV correspond to the focus point region \cite{focuspoint} where the neutralino is mixed gaugino-higgsino \cite{ds-xmssm}.  However, there is a small region near $m_{\N_1}\sim 75$ GeV that has a relic density that is too large, this region is characterized by dominantly Bino neutralinos that annihilate rather weakly.  The Higgs boson pole at $m_{\N_1}\sim 60$ GeV increases the annihilation rate, allowing the relic density to match the observed values.

The annihilations of neutralinos captured by the Sun or Earth may potentially be detected with neutrinos in the SuperKamiokande and IceCube experiments.  Neutralino annihilations that occur on galactic scales may be detectable from measurements of the flux of gamma rays \cite{gammaray}, positrons \cite{positron} or antideuterons \cite{antideuterons}.

\begin{figure}[tb]
\hspace{.7in}
\includegraphics[width=0.35\textwidth,angle=-90]{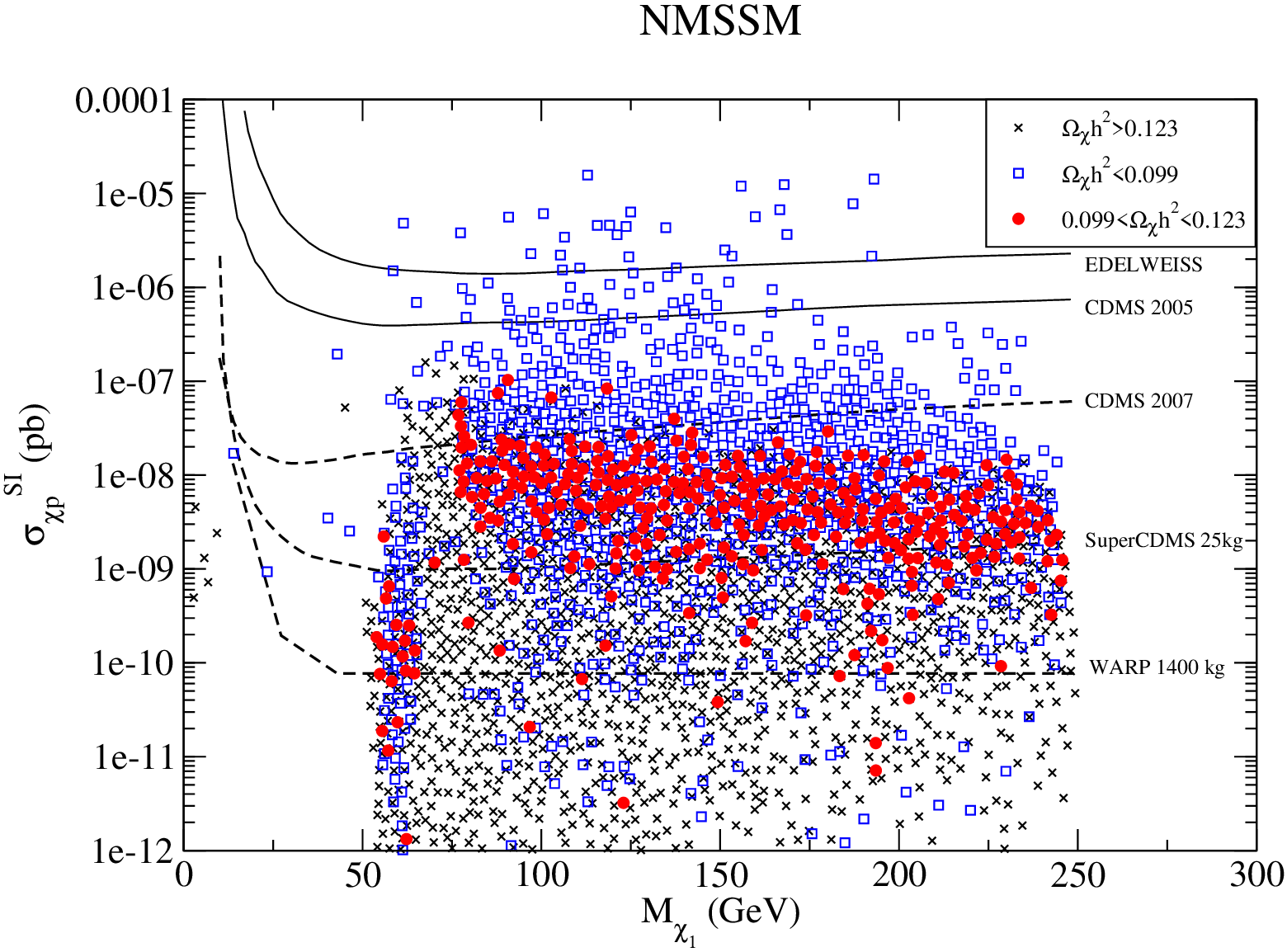}
\includegraphics[width=0.35\textwidth,angle=-90]{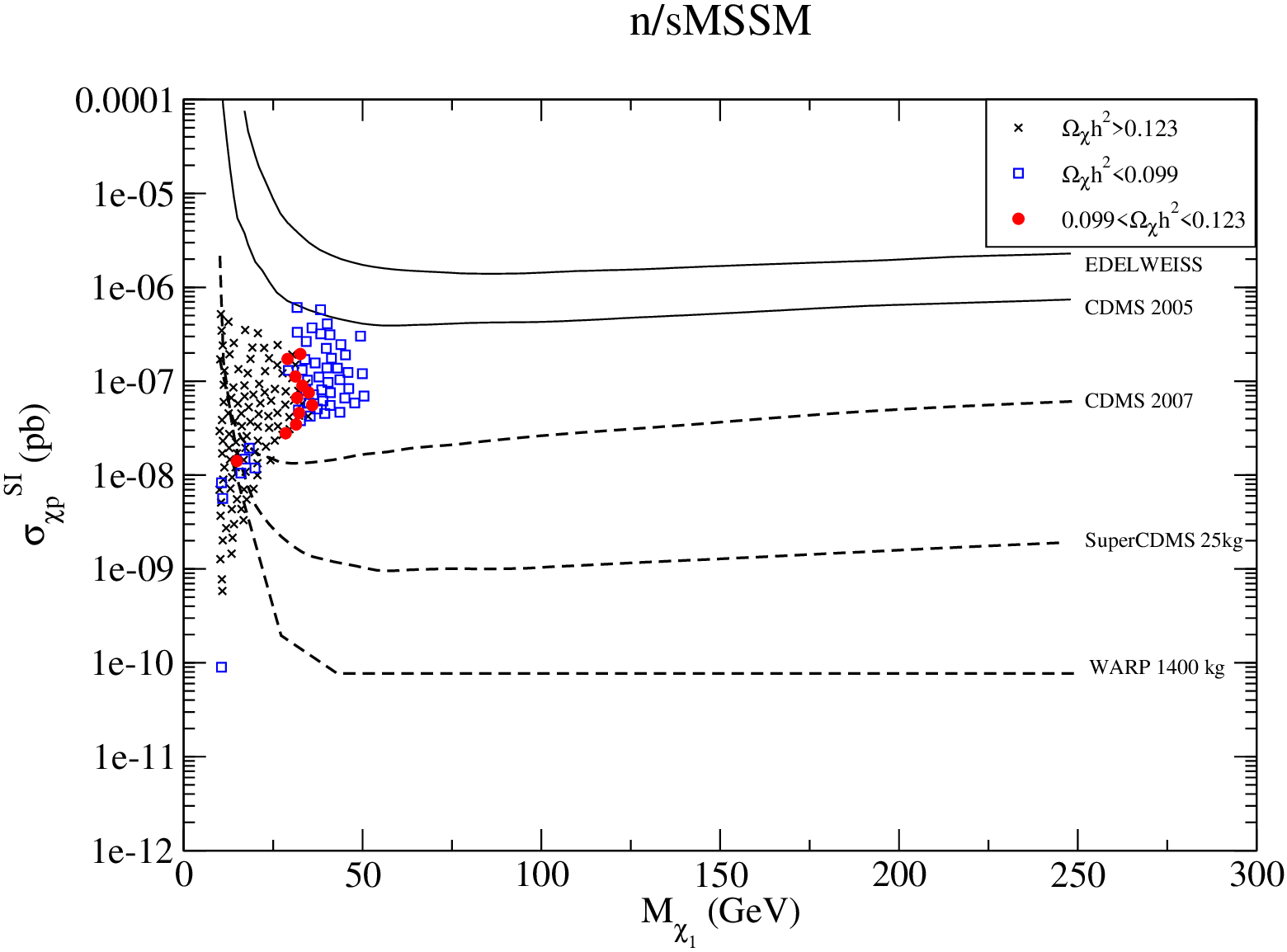}
\caption{Expected spin independent direct detection cross section predictions of the NMSSM and n/sMSSM.  The NMSSM results are similar in distribution to the MSSM and UMSSM.  The expected sensitivities of EDELWEISS, CDMS II (2005), CDMS 2007, SuperCDMS (25 kg) and WARP experiments are shown.  Over most of the neutralino mass range the experiments should detect the signals from the MSSM, NMSSM and UMSSM.  However, if the neutralino annihilates via a Higgs boson resonance, the relic density may be in the preferred region while the direct detection rate is out of reach of future experiments.  Note that the prediction have intrinsic uncertainties of order $\pm60\%$.  Figures from Ref. \cite{ds-xmssm}}
\label{fig:scatt}
\end{figure}

Direct detection experiments look to discover the dark matter particle from the recoil of nuclei from neutralino scattering.   For a review of the dark matter recoil experiments, see Ref. \cite{DMSAG}.  The predictions and current experimental limits for spin-independent scattering cross section off a proton target are shown in Figure \ref{fig:scatt}.

Most of the points in Figure \ref{fig:scatt} where the observed relic density is reproduced are potentially observable by SuperCDMS 25 kg.  However, some points fall below the sensitivity of the WARP experiment \cite{WARP}.  These points are due to annihilation through the Higgs pole where resonance enhancement of the annihilation rate balances the weak $\N_1\N_1 H_i$ coupling to yield the correct relic density.  If no signal is observed at CDMS 2007 \cite{Akerib:2005za}, the n/sMSSM may well be ruled out since the strong limit on the mass by the WMAP observation forces the scattering cross section to be in an observable region.  Exceptions include the possibility of annihilation through a light $A_1$, allowing a lighter neutralino mass.

Another important aspect of singlet extended models concerns the possibility of
electroweak baryogenesis. It is difficult to obtain a sufficiently strong first
order phase transition in the MSSM because there is no cubic term in the tree-level scalar potential,
and the loop-induced effects are only sufficiently strong for a small region of parameter space
involving a light stop~\cite{ewbgmssm}. However, a strong first order transition is considerably easier
to achieve in the singlet extended models because of the tree-level scalar coupling
due to the supersymmetry breaking $A$ term $\lambda A S H_u \cdot H_d$ associated
with the trilinear superpotential term in (\ref{trilinear})~\cite{extendedewbg}.
There may also be additional CP-violating phases associated with the extended Higgs sector.

\subsection{Singlet Higgs conclusions}

Higgs singlet extensions of the MSSM provide well motivated solutions to the $\mu$ problem.  Including an additional Higgs singlet increases the number of CP-even and CP-odd Higgs states and increases the number of associated neutralino states.  The extended models have interesting consequences in collider phenomenology.  Specifically, we find:

\bi

\item The lightest Higgs can be lighter than the LEP limit of $m_h > 114$ GeV due to reduced Higgs couplings to SM fields due to singlet-doublet mixing; the production rates of these Higgs states are often below the rates for the lightest MSSM Higgs boson.  

\item Direct observation of the lightest Higgs at the LHC is favored for the MSSM, NMSSM and UMSSM.  In the n/sMSSM, the traditional discovery modes can be spoiled by the decay to invisible states such as neutralinos.  However, indirect observation of the Higgs can be employed for the n/sMSSM where invisible Higgs decays to neutralino pairs are often dominant.  

\item The extended models can have an approximately decoupled neutralino that is dominantly singlino, accompanied by an approximate MSSM spectrum of neutralino states.  The lightest neutralino is typically very light in the n/sMSSM, often below 50 GeV, and can affect the predicted multiplicity of leptons significantly.  The rate for $\N_{i\ge2} \C_1$ production increases since $\N_i$ is lighter than it would otherwise be in the MSSM.  The decoupled neutralinos in the NMSSM and UMSSM are typically heavy.  

\item Chargino decays are indirectly affected via their decays to a lighter neutralino state.  The number of neutralino states lighter than the chargino and their modified compositions alter the chargino branching fractions.  The chargino can decay to an MSSM-like $\N_2$ and a singlino $\N_1$, yielding a 5 lepton signal.  Additionally, the extra step in a neutralino decay can allow a 7 lepton final state.  

\item Scenarios exist where the singlet extended models are difficult to differentiate from the MSSM using only the Higgs sector.  However, complementary avenues are available through the discovery of a $Z'$ boson in the UMSSM or extended neutralino cascade decays due to the different neutralino spectrum in singlet extended models.

\end{itemize}

\section*{Acknowledgments}
This work was supported in part by the U.S.~Department of Energy under grant No. DE-FG02-95ER40896, by the Wisconsin Alumni Research Foundation, by the Friends of the IAS, and by the National Science Foundation grant No. PHY-0503584.

\end{document}